\documentclass[twocolumn]{aa}
\usepackage{epsfig}
\def\simlt{\ \raise -2.truept\hbox{\rlap{\hbox{$\sim$}}\raise5.truept   %
\hbox{$<$}\ }}
\def\simgt{\ \raise -2.truept\hbox{\rlap{\hbox{$\sim$}}\raise5.truept   %
\hbox{$>$}\ }}                                                          %

\def\be{\begin{equation}}
\def\ee{\end{equation}}
\def\newline{\hfil\break}

\def\la{\mathrel{\hbox{\rlap{\hbox{\lower4pt\hbox{$\sim$}}}\hbox{$<$}}}}
\def\ga{\mathrel{\hbox{\rlap{\hbox{\lower4pt\hbox{$\sim$}}}\hbox{$>$}}}}

\def\bess{{\cal B}}
\def\mug{$\mu$G ~}
\begin{document}

\title{Where does the hard X-ray diffuse emission in clusters of galaxies come from?}

   \author{S. Colafrancesco \inst{1}, P. Marchegiani  \inst{2} and G.C. Perola \inst{1,2}}

   \offprints{S. Colafrancesco}

\institute{   INAF - Osservatorio Astronomico di Roma,
              via Frascati 33, I-00040 Monteporzio, Italy\\
              Email: cola@mporzio.astro.it
 \and
              Dipartimento di Fisica, Universit\`a Roma 3, Via della Vasca Navale 84, Roma, Italy
             }

\date{Received: Sept. 2004 / Accepted: Jun. 2005 }

\authorrunning {S. Colafrancesco et al.}

\titlerunning {Cluster Radio Halos}

\abstract{The surface brightness produced by synchrotron radiation in Clusters of Galaxies with a
radio-halo sets a degenerate constraint on the magnetic field strength, the relativistic electron
density and their spatial distributions, $B(r)$ and $n_{rel}(r)$, in the intracluster medium. Using
the radio-halo in the Coma Cluster as a case-study, with the radio brightness profile and the spectral
index as the only constraints, predictions are made for the brightness profiles expected in the 20-80
keV band due to Inverse Compton Scattering (ICS) by the relativistic electrons on the Cosmic Microwave
Background, for a range of central values of the magnetic field, $B_0$, and models of its radial
dependence, $B(r)$ (of which two represent extreme situations, namely a constant value either of $B$
or of $n_{rel}(r)$, the third a more realistic intermediate case). It is shown that the possible
presence of scalar fluctuations on small scales in the strength of $B$ tends to systematically depress
the electron density required by the radio data, hence to decrease the ICS brightness expected. These
predictions should be useful to evaluate the sensitivity required in future imaging HXR instruments,
in order to obtain direct information on the spatial distribution and content of relativistic
electrons, hence on the magnetic field properties. If compared with the flux in the HXR tail, whose
detection has been claimed in the Coma Cluster (Fusco-Femiano et al. 2004), when interpreted as ICS
from within the radius $R_h$ of the radio-halo, as measured so far, the predictions lead to central
values $B_0$ which are significantly lower than those which have been obtained (albeit still
controversial) from Faraday Rotation measurements. The discrepancy is somewhat reduced if the
radio-halo profile is hypothetically extrapolated out to $R_{vir}$, that is about three times $R_h$,
or, as suggested by hydrodynamical simulations (Dolag et al. 2002), if it is assumed that $B(r)
\propto n_{th}(r)$. To be noted that in the latter case $n_{rel}(r)$ has its minimum value at the
center of the cluster.
 If real and from ICS, the bulk of the HXR tail should then be contributed by electrons
other than those responsible for the bulk of the radio-halo emission. This case illustrates the need
for spatially resolved spectroscopy in the HXR, in order to obtain solid information on the
non-thermal content of Clusters of Galaxies.

}

 \maketitle


\section{Introduction}

Radio halos in galaxy clusters are extended sources with a low surface brightness permeating the
central cluster regions. They show an approximately regular shape, resembling the cluster X-ray halo,
a steep radio spectrum $J_{\nu} \propto \nu^{-\alpha_R}$ with spectral index $\alpha_R$ typically
$\simgt 1$, and little or no polarization (see, e.g., Feretti 2003 for an observational review).
Cluster radio halos are generally quite extended with radii $R_h \simgt 1 $ Mpc, even though smaller
halos (with $R_h \simlt 0.5$ Mpc radii) have also been detected. The one in the Coma cluster, which is
usually considered the archetype of this population of radio sources, has $R_h \approx 0.9
h_{70}^{-1}$ Mpc.

The spectral slopes fall in the range $0.7 \simlt\alpha_R \simlt 2$ (as obtained from WENSS and NVSS
observations at 327 and 1400 GHz, respectively; see, e.g., the data compilation in Kempner \& Sarazin
2001). In a few well studied cases, with a wider frequency coverage, the spectral shape is known with
a better accuracy. In particular, the Coma radio-halo spectrum has been studied from $\sim 30$ MHz to
$\sim 5$ GHz, and can be fitted by a single power-law with $\alpha_R \approx 1.35$ in the frequency
range $30-1400$ MHz (Deiss et al. 1997). More recent data indicate a steepening at $\nu > 1.4$ GHz
(Thierbach et al. 2003). Some evidence has been also reported (Giovannini et al. 1993) that the
spectral index $\alpha_R$ between $\nu = 326$ MHz and $1.4$ GHz changes from $\approx 0.8$ at an
angular distance $\theta \approx 3$ arcmin from the Coma cluster center up to $\approx 1.4$ at $\theta
\approx 12$ arcmin.
In this respect, however, Deiss et al. (1997) pointed out that the emission at 1.4 GHz is much more
extended than found by Giovannini et al. (1993), thus indicating that the radial increase of
$\alpha_R$ might be in fact weaker than claimed.

The widely accepted view is that the radio halos (see, e.g., Brunetti 2003 for a theoretical review)
are produced by synchrotron emission from a population of relativistic electrons with energies $E_e
\approx 7.9 B^{-1/2}_{\mu} (\nu/GHz)^{1/2} {\rm GeV}$ diffusing in the intra-cluster magnetic field
$B$. The electron energy must be $E_e\simgt 1.37 B^{-1/2}_{\mu}$ GeV in order to emit at frequencies
$\nu \simgt 30$ MHz. Here $B_{\mu}$ is the value of the intra-cluster magnetic field in the
synchrotron emission formulae (see, e.g., Longair 1994), given in $\mu$G units.

The radio flux due to a population of relativistic electrons, whose radial density distribution $\eta
(r)$ and energy spectrum is given by
 \be
 n_{rel}(E,r) = n_{rel,0} E^{-x} \eta(r)
 \label{eq.el.spectrum}
 \ee
can be written as
 \be
 F_{\nu,S} \propto n_{rel,0}  \nu^{- \alpha_R} \int dr r^2 \eta (r) [B_{\bot}(r)]^{\alpha_R + 1} \,,
 \label{eq.radioflux}
 \ee
with $\alpha_R = (x-1)/2$. It clearly depends, in a degenerate way, on the combination of two
quantities: the spectrum and the spatial distribution of the relativistic electron density,
$n_{rel}(E,r)$, and the strength and structure of the transverse magnetic field $B_{\bot}(r)$. In
order to resolve this degeneracy, an independent estimate of one of the two quantities is therefore
needed.

As noted long ago (Perola \& Reinhardt 1972, Harris \& Grindlay 1979, Rephaeli 1979), a direct probe
would consist in measuring the emission inevitably produced by the Inverse Compton Scattering (ICS) of
the same relativistic electrons on the Cosmic Microwave Background (CMB) photons. In particular,
relativistic electrons with energies $E_e \simgt$ a few GeV would scatter the CMB photons up to the
X-ray and gamma-ray energy range. The ICS, on the CMB photons, due to the synchrotron emitting
electrons inevitably produces a power-law emission spectrum
 \be
 F_{E,ICS} \propto n_{rel,0} E^{- \alpha_x} \int dr r^2 \eta(r) \,
 \label{eq.hxrflux}
 \ee
with $\alpha_x \equiv \alpha_R$. Thus a measure of this quantity would yield the number of electrons
involved.

By using Coma cluster as a case-study, the goal of this paper is, in the first place, to predict the
brightness distribution of the ICS in the Hard X-Rays (HXR, specifically in the band 20--80 keV) for
different choices on the radial dependence, on the large scale, of the magnetic field, $B(r)$, using
as the only constraint the available information on the (azimuthally averaged) radio brightness
distribution. Such predictions are not based on the theoretical issues concerning the origin of
non-thermal phenomena in galaxy clusters (for a review of this topic see, e.g., Brunetti 2003), hence
in this respect can be considered as model-independent. We have particularly in mind that our
predictions might be useful to assess the potential of HXR imaging devices in this field. In this
context, the consequences on the ICS emission related to the small scale structure of the magnetic
field, both in direction and strength, will be emphasized.

In the second place, and to its natural extent for what concerns the small scale structures of the
magnetic field just mentioned, the information available on its longitudinal strength, as it can be
obtained, and has been inferred already, from Faraday Rotation (FR) measurements, will be used to make
what at present can be regarded as our ``best'' prediction on the spatially integrated ICS HXR
emission. The latter will then be confronted with observational results, which claim that this
component has been already detected (Fusco-Femiano et al. 1999, 2004, Rephaeli et al. 1999).

The plan of the paper is the following. In Sect.2 the radio brightness profile in Coma is used to
constrain, in a parametric form, the radial dependence of the relativistic electron density and that
of the magnetic field, and the relationships between the two. Sect. 3 is devoted to the impact of both
scalar and vectorial fluctuations in the magnetic field on the synchrotron emissivity. In Sect. 4 the
relativistic electron density as a function of radius is derived for a set of representative values of
the central magnetic field, $B_0$, and three different radial profiles for $B(r)$. Correspondingly, in
Sect. 5 the predicted ICS brightness profiles are illustrated. In Sect. 6 the results are confronted
with the measurement of a HXR candidate ICS spectral tail in Coma, and discussed also in the light of
the available information on the strength of $B$ from Faraday Rotation measurements. Sect. 7 contains
summary and conclusions.  We use $H_0=70 km\, s^{-1} Mpc^{-1}$ and a flat, vacuum-dominated cosmology
($\Omega_{\Lambda}=0.7, \Omega_m \approx 0.3$) throughout the paper.

\section{The radio-halo surface brightness in Coma}

We derive here the constraints on the spatial structure of the magnetic field $B(r)$ and of the
relativistic electrons density, $n_{rel}(E,r)$, in the Coma cluster from the analysis of its
radio-halo surface brightness. Since the radio halo emission depends on both the magnetic field
amplitude and the relativistic electron density in a degenerate way (see eq. \ref{eq.radioflux}), the
magnetic field spatial structure will be inferred from the radio halo surface brightness observations
by assuming parametric models for the spatial distribution of the relativistic electrons.

First of all, we fit the azimuthally averaged radio brightness profile of Coma observed at $\nu= 1.4$
GHz (Deiss et al. 1997) assuming initially the profile
\begin{equation}
S_\nu(\theta)=S_{\nu,0} \bigg[1+ \bigg({\theta \over \theta_c} \bigg)^2 \bigg]^{-q_R'}
 \label{bril.inf}
\end{equation}
and using the same core radius which fits the cluster X-ray brightness, with $\theta_c=10.5'$ (Briel
et al. 1992). In this case, we find a best fit parameter $q_R'=1.63 \pm 0.09$ with a reduced $\chi^2$,
$\chi^2_{red}=2.13$ with 6 d.o.f. (we give hereafter $1 \sigma$ uncertainties).

\begin{figure}[htbp]
\begin{center}
 \epsfig{file=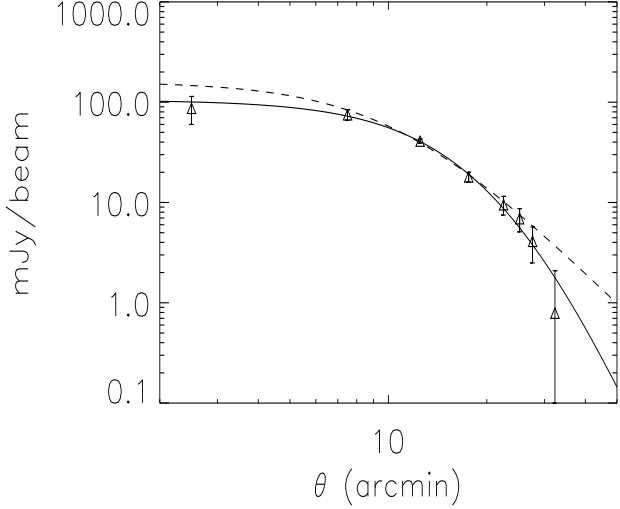,height=8.cm,width=9.cm,angle=0.0}
  \caption{\footnotesize{The spatial distribution of the radio-halo brightness
  of the Coma cluster fitted with the functional form in eq. (\ref{bril.inf}): the dashed curve represents
the fit with $\theta_{c}$ fixed at 10.5 arcmin form the X-ray profile, the solid curve with the best
fit value of $\theta_{c,R}=23.8$ arcmin. Data are from Deiss et al. (1997).
  }}\label{fig.fitbril}
\end{center}
\end{figure}

Because this fit is not sufficiently accurate, as can be judged in Fig.\ref{fig.fitbril}, it was
repeated letting this time both $q_R'$ and $\theta_{c,R}$ as free parameters. In this case a much
better fit is obtained ($\chi^2_{red}=0.40$ with 5 d.o.f., see Fig.\ref{fig.fitbril}), with $q_R'=
3.9\pm1.2$ and $\theta_{c,R}= 23.8\pm6.2$ arcmin. In the following, the parameter values from the
second fit will be used.

By adopting these values, along with a value of the spectral index independent from $\theta$, we are
implicitely assuming that they apply also to the emission at frequencies lower than $\nu=1.4$ GHz. At
this frequency the electron energy is related to the field strength as $E_e \approx 9.35
B^{-1/2}_{\mu}$ GeV. Electrons with this energy emit via ICS photons with energy $E_X\sim 700
B_\mu^{-1}$ keV. Therefore, for $B<8.75\, \mu$G, the 20-80 keV ICS emission is due to electrons which
emit in radio at frequencies $\nu \simlt (350-1400)$ MHz. Since we shall confine ourselves to $B$
values smaller than the limit above, the spectral steepening observed at $\nu > 1400$ MHz (Thierbach
et al. 2003) has no effects on our predictions.

On the other hand, in order to emit via ICS in the 20--80 keV band, the electrons must have energies
$E_e \approx 0.35 GeV (E/keV)^{1/2}$ $ \approx 1.6-3.2$ GeV. The radio frequency at which these
electrons emit via synchrotron is in the range $\nu \approx (40-60) B_{\mu}$ MHz. For a value of $B$
as small as $B$ = 0.2 $\mu$G, their radio emission would fall in the range $\approx 8-32$ MHz, which
lies completely outside the frequency span of the observations available (see, e.g., Deiss et al.
1997). Therefore, one must be aware that any ICS prediction for values of $B_{\mu}$ less than unity
relies, to an extent which increases when $B$ decreases, upon a simple extrapolation of the observed
spectrum.

The radial dependence of the synchrotron emissivity, which corresponds to the brightness profile
assuming spherical symmetry, is given by
\begin{equation}
J(r) \propto [1+(r/r_{c,R})^2]^{-q_R}
 \label{eq.radiobrightnessprofile}
\end{equation}
with best-fit parameters $q_R = q_R'+1/2 = 4.4$ and $r_{c,R} = 0.67h_{70}^{-1}$ Mpc.

To convert this dependence into constraints on $B(r)$ and $n_{rel}(r)$, we adopt a simple functional
form, namely $n_{rel}(E,r)= n_{rel,0} E^{-x} \eta(r)$ (as in eq.\ref{eq.el.spectrum}) $B(r) = B_0
g(r)$, where
\begin{eqnarray}
\eta(r) &=& [1+(r/r_{c,R})^2]^{-q_e}\\
   g(r) &=& [1+(r/r_{c,R})^2]^{-q_B},
   \label{eq.radialprofiles}
\end{eqnarray}
with the same core radius $r_{c,R}$. It follows that
\begin{equation}
 J(r) \propto \eta(r) g(r)^{\beta_R}
 \label{eq.radiobrightnessprofile2}
 \end{equation}
with $\beta_R=\alpha_R+1=2.35$ (with the adopted value $\alpha_R = 1.35$). In order to reproduce the
observed brightness profile, the following relationship must be obeyed:
\begin{equation}
 q_e+\beta_R q_B = q_R
 \label{eq.qvalues}
 \end{equation}

We shall consider, for illustration of the HXR brightness profile to be expected, three cases: {\it
i)} a uniform distribution of the relativistic electrons, $q_e=0$, with the corresponding slope
$q_B=q_R/\beta_R=1.87$; {\it ii)} the case of a constant pressure ratio between the magnetic field and
relativistic electrons, $q_e=2q_B$, hence $q_B=1.01$ and $q_e=2.02$; {\it iii)} a uniform magnetic
field, $q_B=0$, with the corresponding slope $q_e=4.4$.

The first and the last should be regarded as limiting cases. We are fully aware of the fact that
according to hydrodynamical simulations (see, e.g., Gon\c{c}alves \& Fria\c{c}a 1999, Dolag et al.
2001, 2002) the magnetic field strength is likely to decrease radially. It is also important to stress
that these simulations indicate, in addition, that the $B$ field fluctuates on a wide range of scale.
These fluctuations can be very relevant when it comes to infer the amount of radio-emitting electrons,
hence of their ICS emission, as we shall see in the next Section.

\begin{figure}[htbp]
\begin{center}
 \epsfig{file=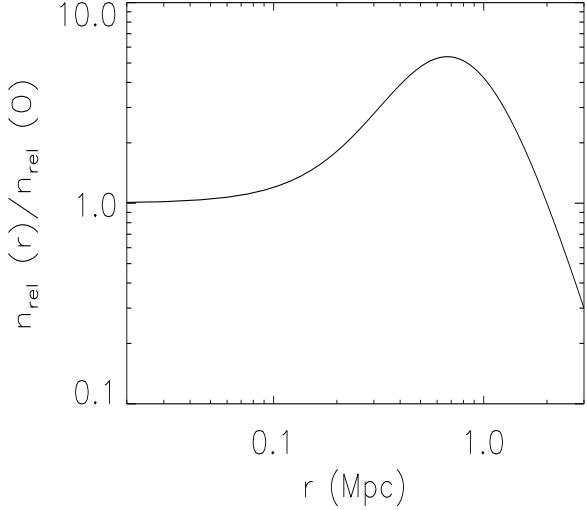,height=8.cm,width=9.cm,angle=0.0}
  \caption{\footnotesize{The radial distribution of the relatvistic electrons,
if $B(r)$ is assumed to go like the density of the thermal intracluster gas, as from hydrodynamical
simulations by Dolag et al. (2002).
  }}\label{fig.dolag}
\end{center}
\end{figure}

Here it is worth stressing that a result from hydrodynamical simulations, which seems to be partly
supported by observational results on Faraday Rotation (Dolag et al. 2002) -- namely that $B(r)$ goes
approximately like the density of the thermal gas -- if assumed to hold for Coma, implies what is
shown in Fig. \ref{fig.dolag}: this figure shows, in fact, that the density of the relativistic
electrons, in order to reproduce the radio brightness profile, would have to increase from the center
outward up to a maximum at about 0.8R$_h$. We regard this as a rather unlikely situation, a
consequence which seems, to our knowledge, to have been disregarded so far. We shall however return to
this situation in Sect. 6.

\section{The role of magnetic field structure: radial profile and  fluctuations}

The intra-cluster magnetic field is likely consisting of a smooth component on top of which there are
scalar (amplitude) and vectorial (directional) fluctuations.

Because the synchrotron emissivity depends non-linearly on the amplitude of the transverse component
$B_\bot$ of the magnetic field, $J_\nu \propto B_{\bot}^{\alpha_R+1}$, a radial dependence in the
smooth component of $B$, as well as the fluctuations, do have an impact on the number of relativistic
electrons responsible for the same radio-halo flux.

Concerning the smooth component with profile $B(r)=B_0 g(r)$, one can introduce its volume-averaged
value:
\begin{equation}
 <B> = {\int dV B(r) \over \int dV}= 3 B_0 \left(\frac{r_{c,R}}{ R}\right)^3 I \, ,
 \label{eq.averageb}
 \end{equation}
where $I = \int_0^p dx x^2 g(x)$, $x \equiv (r/r_{c,R})$ in terms of the typical core radius $r_{c,R}$
of the magnetic field, and of the maximum radial extension of the magnetic field $R=p r_{c,R}$. For
the same volume-averaged value of the magnetic field, we have verified that the volume integrated
number of relativistic electrons is larger when $q_B = 0$ than for any other value in the interval
$0-1.87$ selected in the previous Section.

To obtain a quantitative estimate of the impact of scalar fluctuations, we adopt a simple treatment.
Let us assume that, on scales smaller than $l(r)$
\begin{equation}
l(r) \ll {dr \over dB } B(r)
\end{equation}
where $B$ can be considered independent of $r$, the strength is subject to fluctuations $\delta B$,
such that
\begin{equation}
\langle B + \delta B \rangle = B
\end{equation}
Hence within $l(r)$
\begin{equation}
\langle (B+\delta B)^2 \rangle = B^2 + \langle (\delta B)^2 \rangle
\label{eq.fluct}
\end{equation}

Upon substitution of eq.(\ref{eq.fluct}) in the synchrotron formula, one obtains
\begin{equation}
J(r) \propto n_{rel}(r) B(r)^{\alpha_R+1} \bigg[ 1+ {\langle (\delta B)^2 \rangle  \over B^2}
\bigg]^{(\alpha_R+1)/2} \label{eq.emissivityfluct}
\end{equation}

The factor containing the second order effect of the fluctuations in eq.(\ref{eq.emissivityfluct}) is
always $\geq 1$, and can itself be a function of $r$. If we assume, for the sake of simplicity, that
it is independent of $r$, then one can immediately compute the factor, after integrating over the
volume to obtain the flux $F_{\nu}$, by which the quantity $n_{0,rel}$ is overestimated with respect
to the case of zero fluctuations. For example, with the value of $\alpha_R$ adopted for Coma, this
factor is equal to $1.12, 1.6, 2.26$  for $\langle (\delta B)^2 \rangle / B^2 = 0.1, 0.5$ and $1$,
respectively.

Vectorial fluctuations are of great importance in the interpretation of Faraday Rotation measurements,
yielding
\begin{equation}
RM=\int n_{th} \vec{B} \cdot d\vec{\ell}
\end{equation}
where $n_{th}$ is the density of thermal electrons inferred from X-ray observations.

These measurements, given the independent constraint on $n_{th}$, are in principle a very good tool to
estimate the strength of the $B$ field parallel to the line of sight, along with the coherence scale
of the field direction. We shall return on this issue in Sect. 6. Here it is sufficient to recall that
the average value of this scale is very much smaller than the cluster radius. To quantify the effects
of the directional fluctuations on the emissivity, we shall assume a random and isotropic
distribution. The emissivity formula will therefore contain the term
\begin{equation}
\langle B_\bot^{\alpha_R + 1}\rangle = f(\alpha_R)B^{\alpha_R +1} = 0.5 \bigg(\int_0^\pi (\sin
\theta)^{2+\alpha_R}d\theta \bigg) B^{\alpha_R +1} \label{eq.random}
\end{equation}

\section{Constraints on the relativistic electron density}

If we use the smooth radial distribution of the intra-cluster magnetic field as given by
eq.(\ref{eq.radialprofiles}), the normalization of the relativistic electron distribution in
eq.(\ref{eq.el.spectrum}) which reproduces the radio-halo flux at 1.4 GHz, $J_{1.4}= 0.64$ Jy (Deiss
et al. 1997), can be written as
 \be
 n_{rel,0} = K_{\alpha_R} I_R^{-1} f^{-1}(\alpha_R)\bigg( \frac{B_{0}}{\mu G} \bigg)^{-(\alpha_R+1)} \,
 cm^{-3} GeV^{-1}
 \label{eq.nrel0}
 \ee
where $I_R = \int_0^1 dx x^2  [ 1 + (R_h/r_{c,R})^2 x^2]^{-q_R}$, $f^{-1}(\alpha_R)$ is given by
eq.(\ref{eq.random}), and, we remind, $q_R = q_e + \beta_R q_B$. For the choice $\alpha_R$ = 1.35, the
normalization factor $K\cdot f^{-1}$ in eq.(\ref{eq.nrel0}) is equal to 8.64$\cdot 10^{-14}$.

For the predictive purposes declared in Sect. 1, we have adopted the following values of the central
magnetic field strength, $B_{0}$ = 0.1, 0.2, 0.5, 1, 3, 5 $\mu$G. For the three combination of $q_e$
and $q_B$ anticipated in Sect. 2, and $\alpha_R=1.35$, the quantity  $n_{rel,0}$ is given in
Tab.\ref{tab.parametri}, Col. 5, along with, in Col. 4, the volume average of $B$ calculated,
according to eq. (\ref{eq.averageb}), {\it within the observed value $R_h$ of the radio--halo}.

\begin{figure}[tbp]
\begin{center}
\vbox{
 \epsfig{file=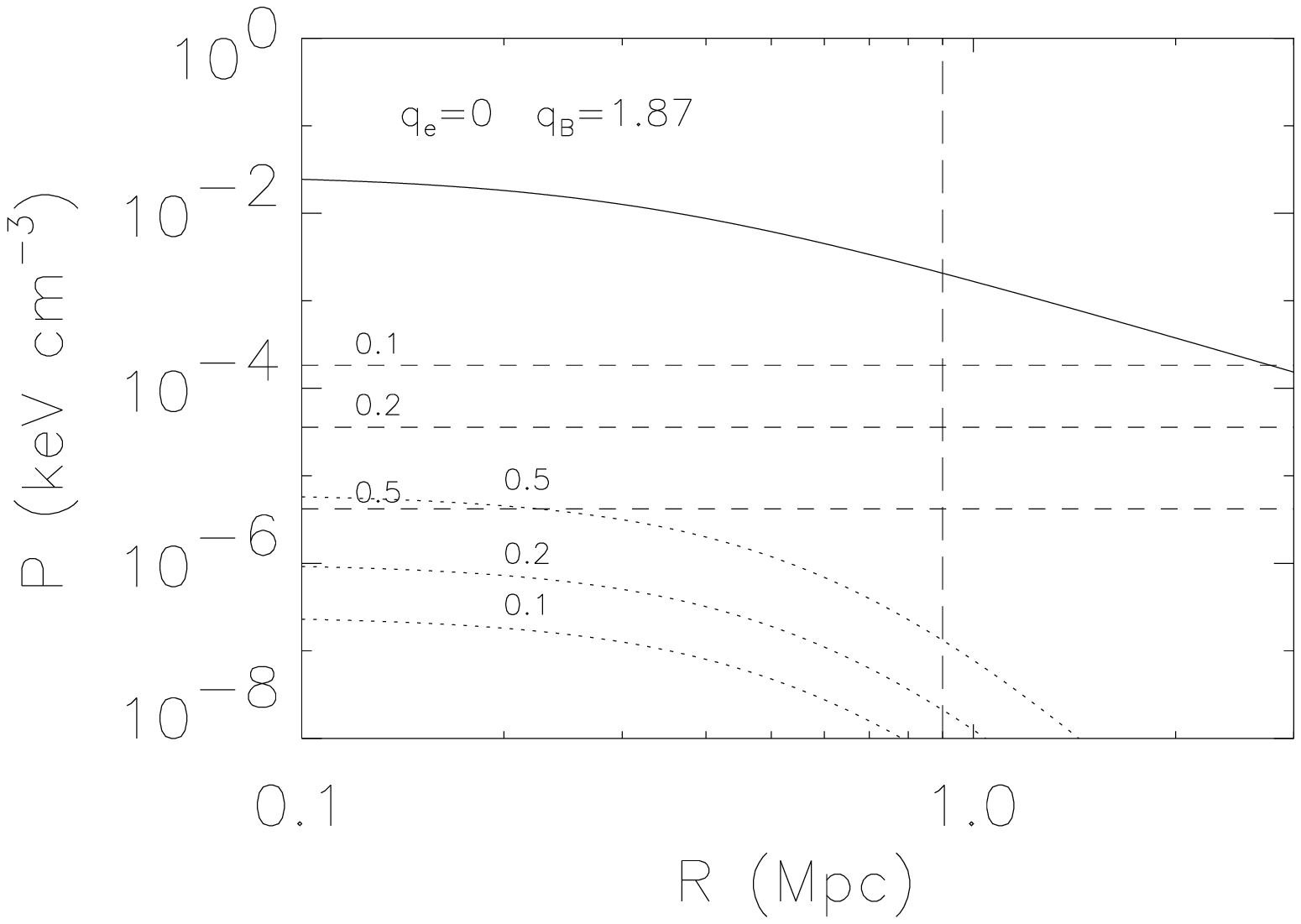,height=8.cm,width=9.cm,angle=0.0}
 \epsfig{file=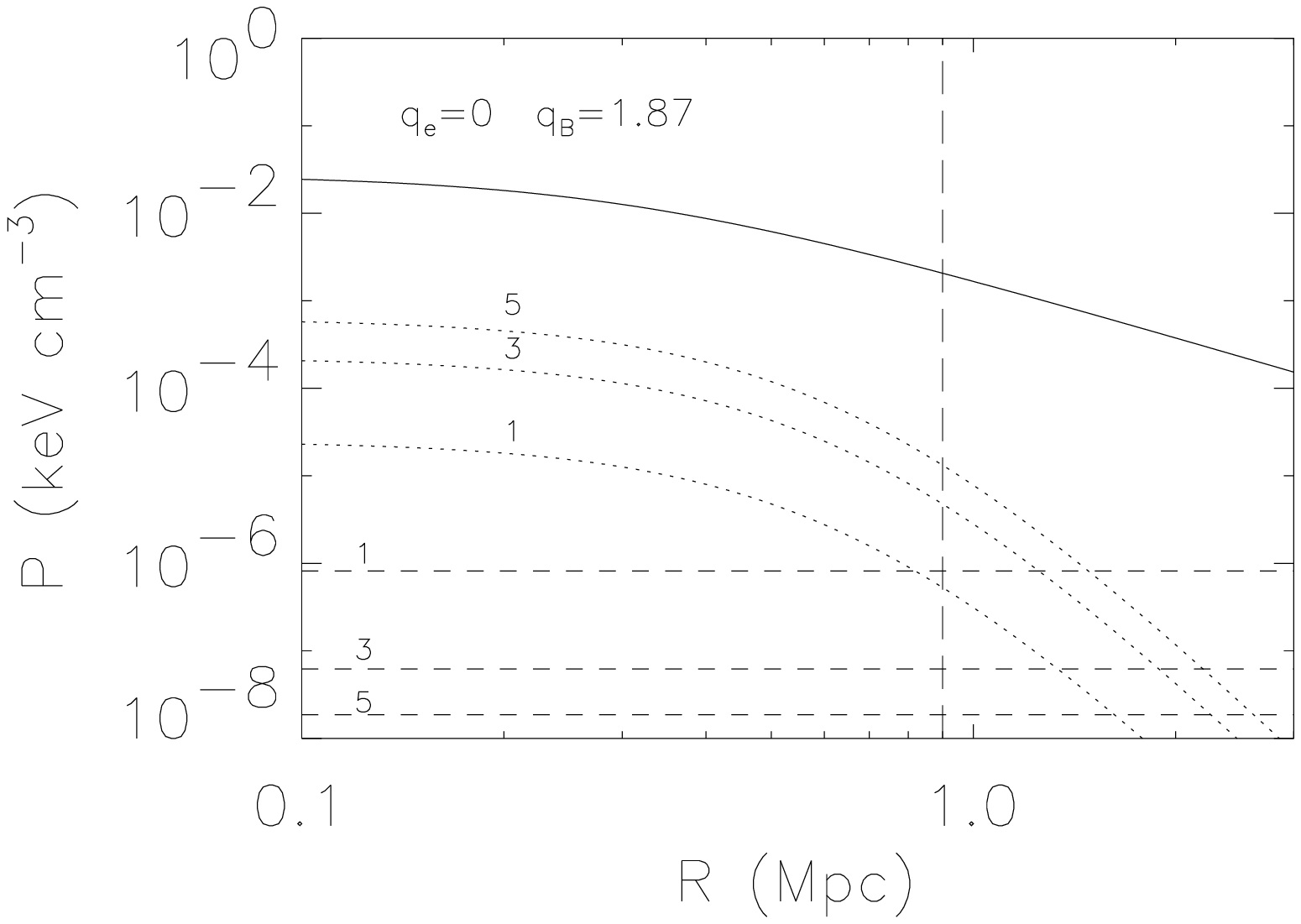,height=8.cm,width=9.cm,angle=0.0}
}
  \caption{\footnotesize{The radial distribution of the pressure of
  the relativistic electrons (dashed) and the magnetic field (dotted) which fit the Coma radio-halo
  brightness. We show the predictions for models with $q_e=0$ and $q_B = 1.87$ for values $B_{ 0} = 0.1, 0.2$
  and $0.5$ $\mu$G (upper panel) and for $B_{0} = 1, 3$ and $5$ $\mu$G (lower panel).
  We also show the pressure of the thermal intra-cluster gas (solid) for comparison. The
  vertical dashed line is placed at $R_h$ = 0.9 Mpc.
  }}\label{fig.pressionia}
\end{center}
\end{figure}

\begin{figure}[tbp]
\begin{center}
\vbox{
 \epsfig{file=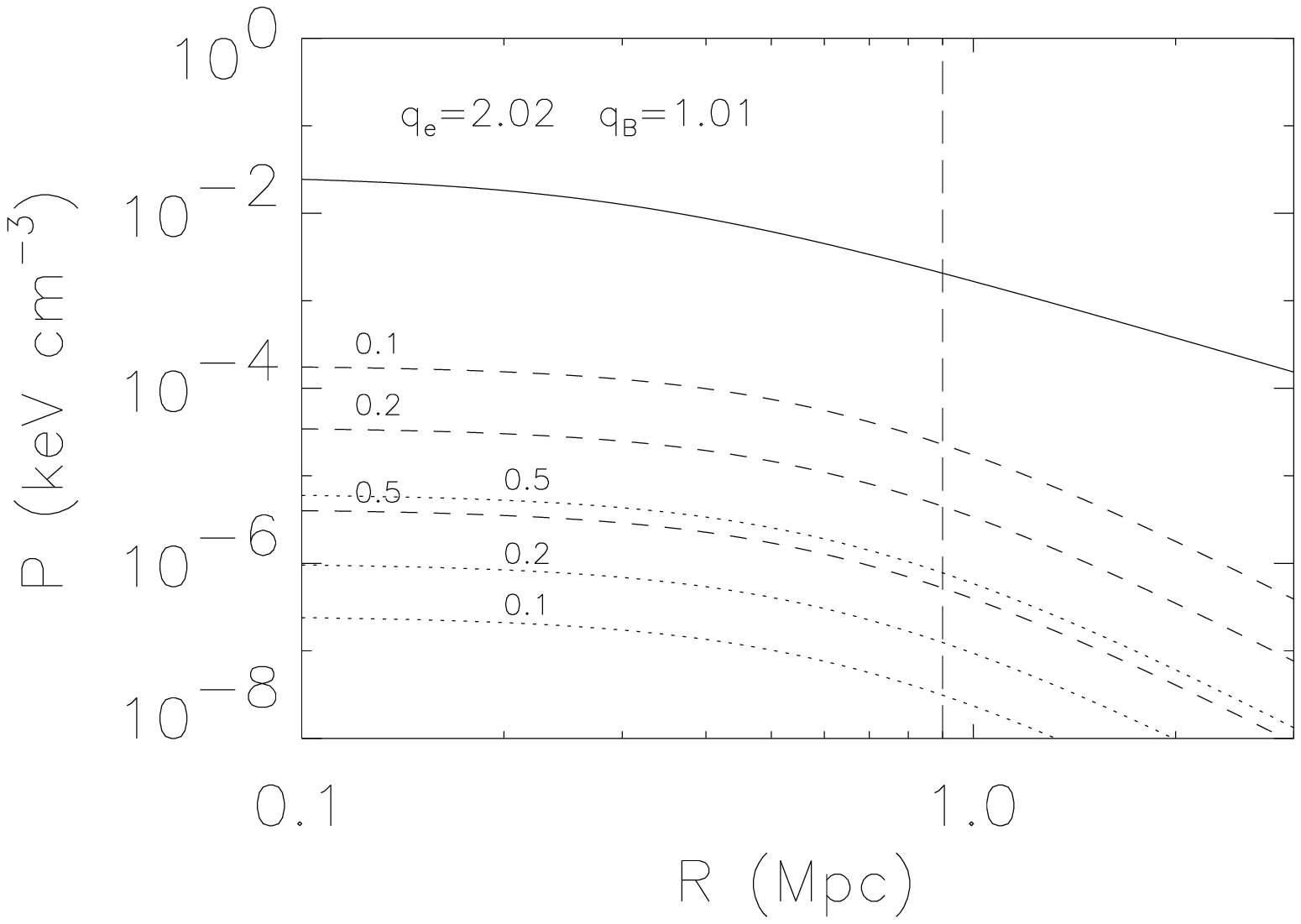,height=8.cm,width=9.cm,angle=0.0}
 \epsfig{file=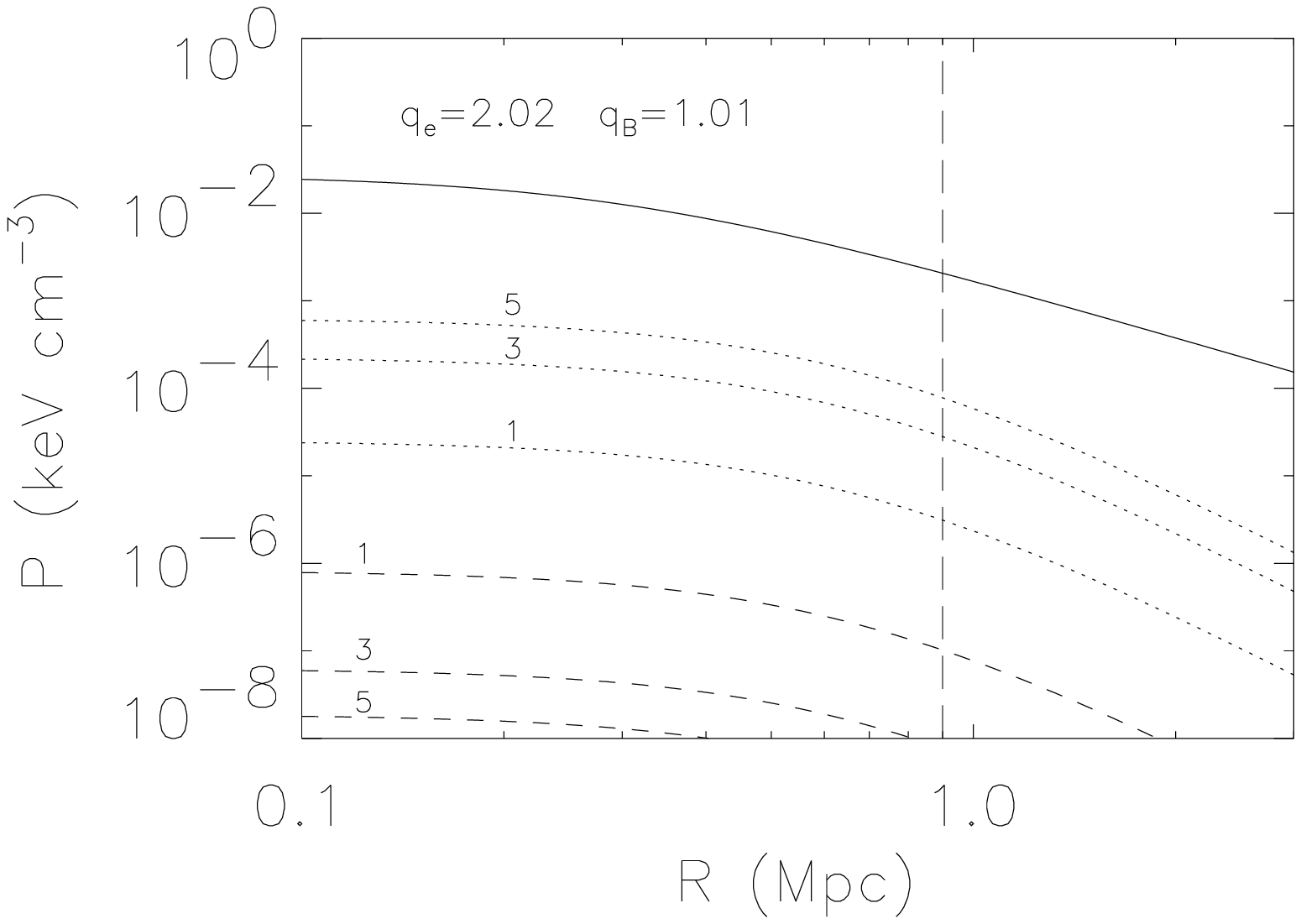,height=8.cm,width=9.cm,angle=0.0}
}
  \caption{\footnotesize{Same as Fig.\ref{fig.pressionia} but for models with $q_e=2.02$ and
  $q_B = 1.01$.
  }}\label{fig.pressionib}
\end{center}
\end{figure}

\begin{figure}[tbp]
\begin{center}
\vbox{
 \epsfig{file=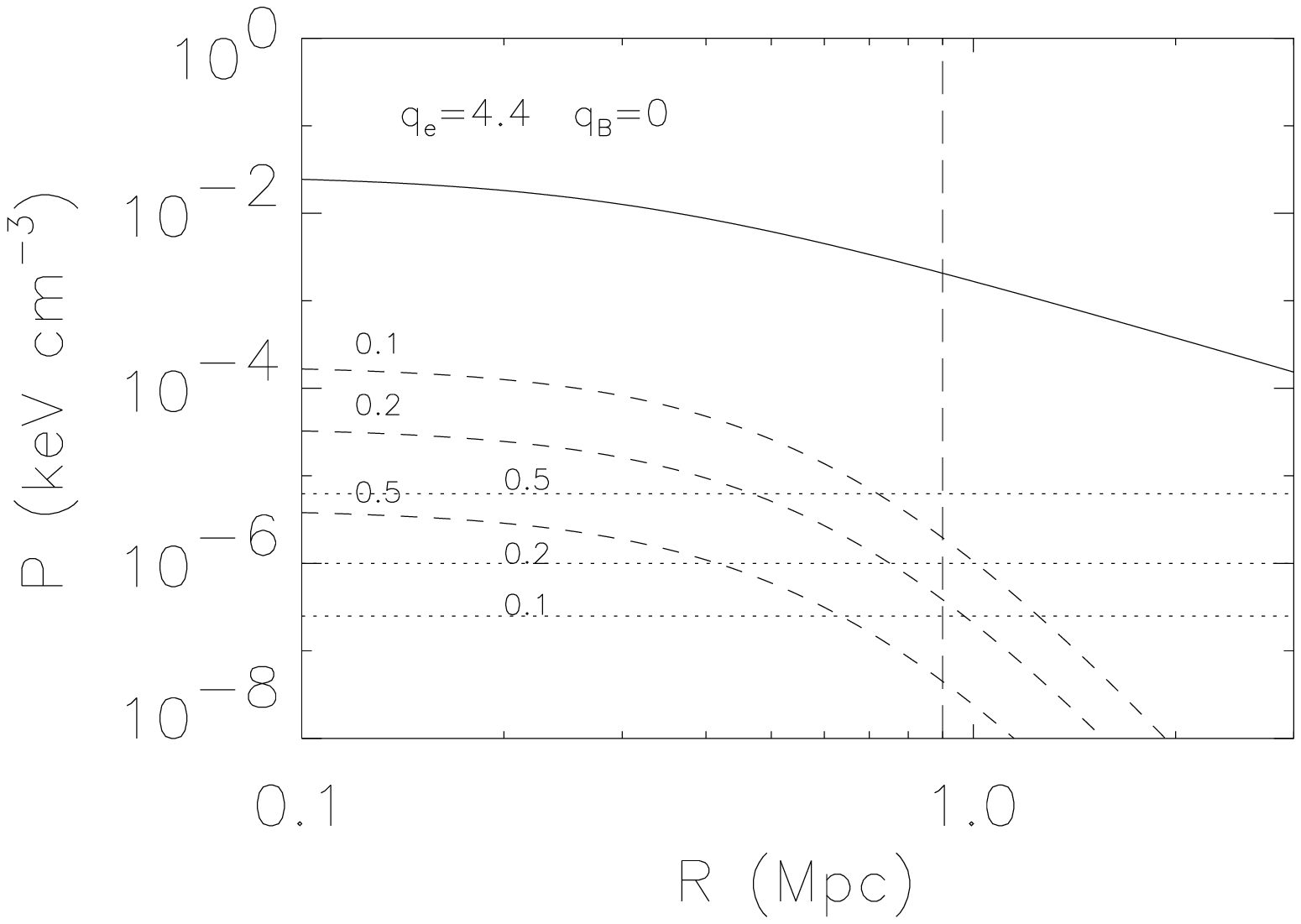,height=8.cm,width=9.cm,angle=0.0}
 \epsfig{file=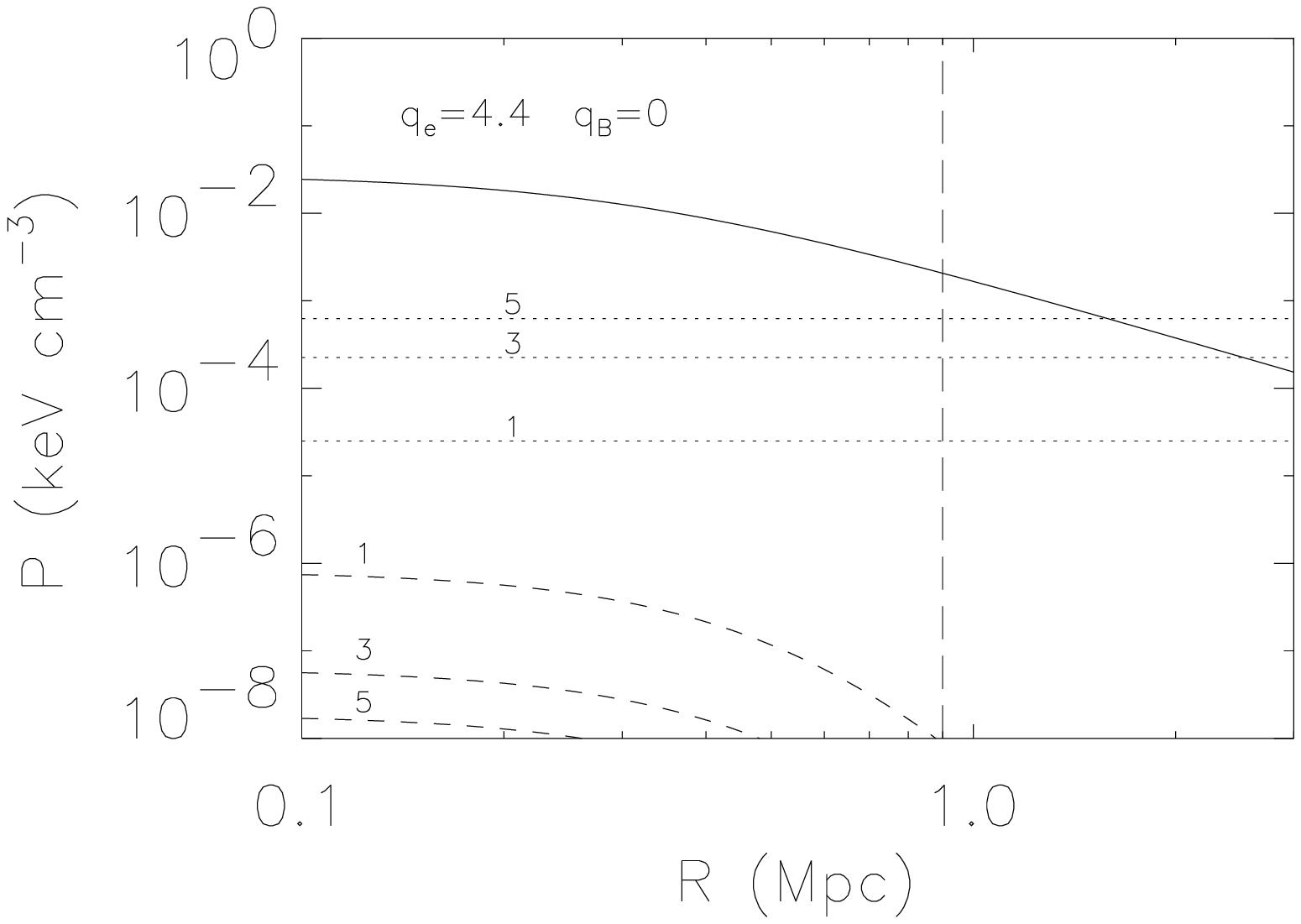,height=8.cm,width=9.cm,angle=0.0}
}
  \caption{\footnotesize{Same as Fig.\ref{fig.pressionia} but for models with $q_e=4.4$ and $q_B = 0$.
  }}\label{fig.pressionic}
\end{center}
\end{figure}

For all the combinations of the selected values of $B_0$, $q_e$ and $q_B$, we show in
Figs.\ref{fig.pressionia}-\ref{fig.pressionic}, as a function of the radial distance from the center,
the pressures contributed by the relativistic electrons, $P_{rel}$, by the magnetic field, $P_B$, and
by the thermal gas, $P_{thermal}$. The relativistic electron pressure was evaluated assuming
$E_{min}=1$ GeV. The pressure profiles in these figures are extrapolated out to $R_{vir}$, which is
approximately equal to 3 Mpc (Lokas \& Mamon 2003).

The pressure of the relativistic electrons needed to fit the radio-halo brightness of Coma strongly
decreases with increasing values of the central magnetic field $B_0$, consistently with
eq.(\ref{eq.nrel0}). More important, it is apparent that the pressure provided by the relativistic
electrons is always smaller than the pressure of the thermal gas. Whithin $R_h$ this holds in all
cases also when the magnetic pressure is added to that of the electrons.

To be kept in mind, as emphasized in Sect. 3, the presence of scalar fluctuations in the magnetic
field $B$ would decrease the values of $n_{rel}$, given in Table 1, by the factor $[ 1+ {\langle
(\delta B)^2 \rangle / B^2}]^{(\alpha_R+1)/2}$.


\section{Inverse Compton Scattering emission}

In this Section we use the radial profiles of the relativistic electrons and of the magnetic field, in
any of the abovementioned combinations which fit the radio halo brightness distribution, to predict
the brightness distribution of the HXR emission produced by ICS of the relativistic electrons on the
CMB photons.

For the general case of a relativistic electrons distribution as in eq. (\ref{eq.el.spectrum}), it is
imperative, in spherical symmetry, to introduce a maximum radius out to which it is assumed to hold,
and beyond which it goes to zero. This is so because the density of CMB photons is a constant, unlike
the strength of the magnetic field, which is constrained, for any choice of $q_e$, by the radio
brightness profile. Here we assume this radius equal to $R_{vir}$ = 3 Mpc, corresponding to a maximum
value of the angular distance from the centre $\Theta = 105$ arcmin.

Among the cases selected, it is evident from Figures \ref{fig.pressionia}, \ref{fig.pressionib} and \ref{fig.pressionic}
 that for some values of $B_0$, when either
$q_e$ or $q_B$ are equal to zero, the non-thermal pressure does not always remain below the thermal
one when one moves beyond $R_h$, and this need to be kept in mind. For simplicity we shall provide the
ICS brightness predictions {\it in all cases} using the same value of $\Theta$.

\begin{figure}[tbp]
\begin{center}
 \epsfig{file=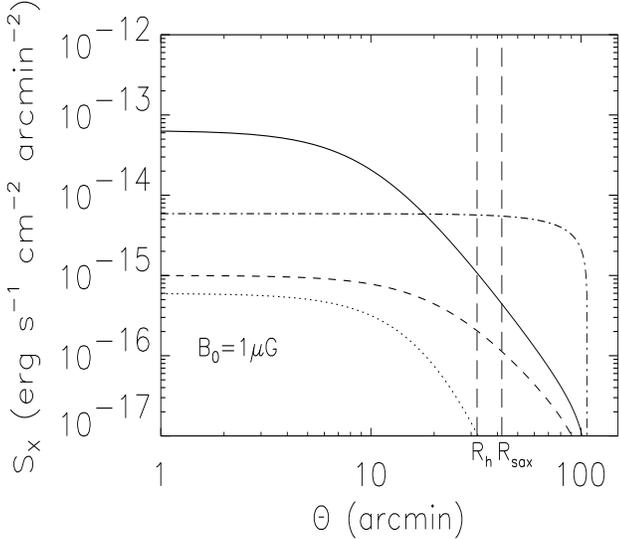,height=8.cm,width=9.cm,angle=0.0}
  \caption{\footnotesize{The spatial distribution of the HXR brightness of the Coma cluster in the 20-80 keV
  energy band as produced by ICS of relativistic electrons reproducing the radio halo flux.
  We show the predictions for different models:
  $q_e=0$ (dot-dashed curve), $q_e=2.02$ (dashed curve) and $q_e=4.4$ (dotted curve)
  for a value $B_{0}=1$ \mug. The thermal bremsstrahlung brightness of Coma integrated in the same energy band
  (solid curve) is also shown for comparison.
  }}\label{fig.brilhxr2}
\end{center}
\end{figure}

\begin{figure}[tbp]
\begin{center}
 \epsfig{file=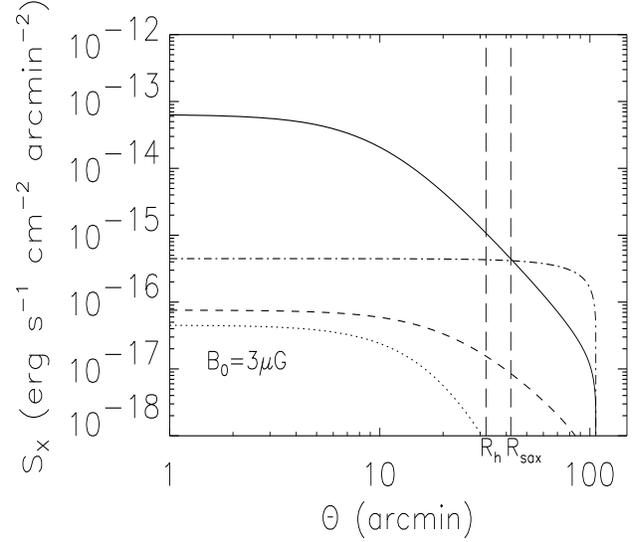,height=8.cm,width=9.cm,angle=0.0}
  \caption{\footnotesize{Same as Fig. \ref{fig.brilhxr2} but for $B_{0}=3$ \mug.
  }}\label{fig.brilhxr3}
\end{center}
\end{figure}

\begin{figure}[tbp]
\begin{center}
 \epsfig{file=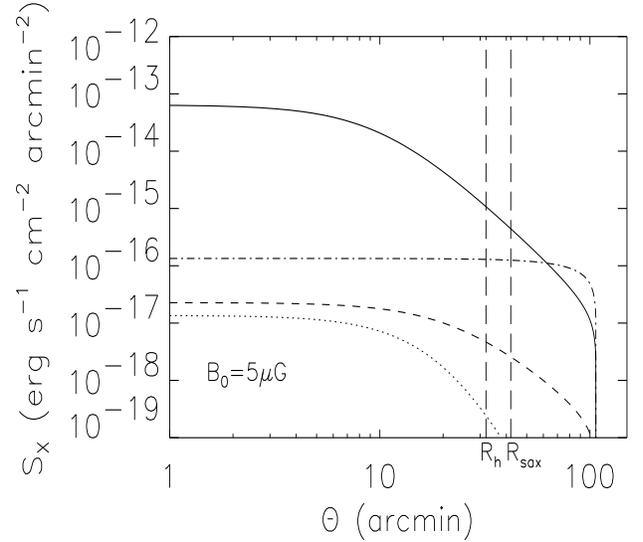,height=8.cm,width=9.cm,angle=0.0}
  \caption{\footnotesize{Same as Fig. \ref{fig.brilhxr2} but for $B_{0}=5$ \mug.
  }}\label{fig.brilhxr4}
\end{center}
\end{figure}

\begin{figure}[tbp]
\begin{center}
 \epsfig{file=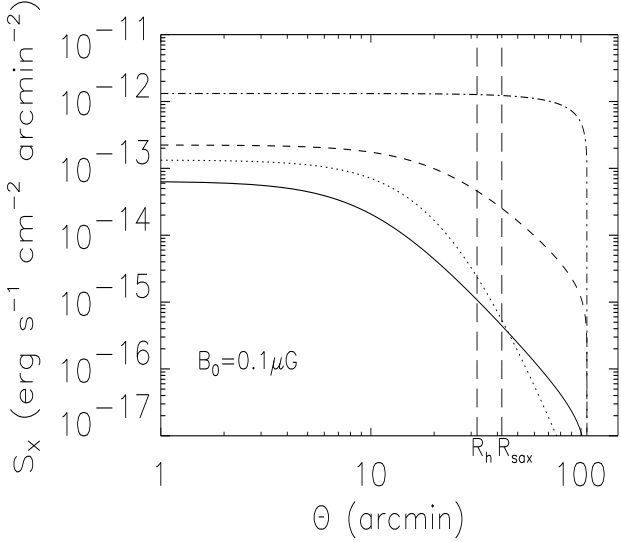,height=8.cm,width=9.cm,angle=0.0}
  \caption{\footnotesize{Same as Fig. \ref{fig.brilhxr2} but for $B_{0}=0.1$ \mug.
  }}\label{fig.brilhxr01}
\end{center}
\end{figure}

\begin{figure}[tbp]
\begin{center}
 \epsfig{file=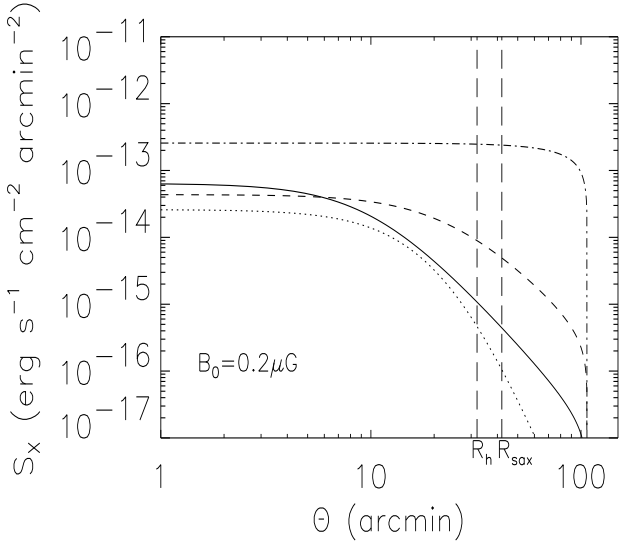,height=8.cm,width=9.cm,angle=0.0}
  \caption{\footnotesize{Same as Fig. \ref{fig.brilhxr2} but for $B_{0}=0.2$ \mug.
  }}\label{fig.brilhxr02}
\end{center}
\end{figure}

\begin{figure}[tbp]
\begin{center}
 \epsfig{file=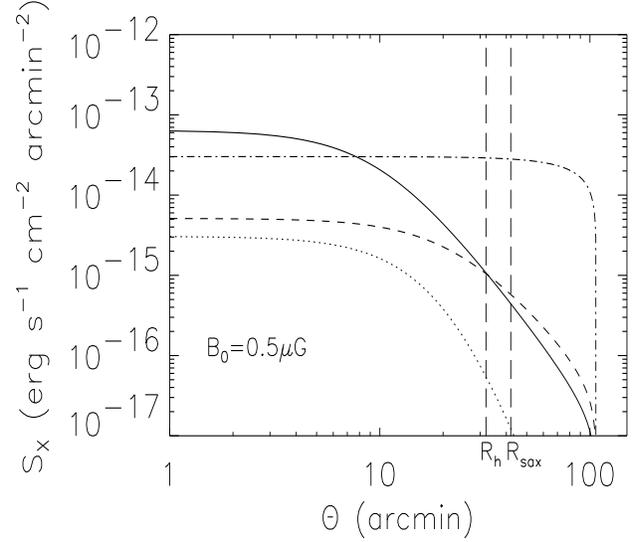,height=8.cm,width=9.cm,angle=0.0}
  \caption{\footnotesize{Same as Fig. \ref{fig.brilhxr2} but for $B_{0}=0.5$ \mug.
  }}\label{fig.brilhxr05}
\end{center}
\end{figure}

The HXR ICS surface brightness at the energy $E_X$ and at the angular distance $\theta$ from the
cluster center writes as
\begin{eqnarray}
S_X(E_X,\theta) & = & S_{X,0} \left [1+(\theta/\theta_{c,R})^2 \right]^{-q_e+1/2}
\label{eq.brilhxr}\\
                & \cdot & [\bess(q_e-1/2,1/2)-\bess_m(q_e-1/2,1/2)]
                \nonumber
\end{eqnarray}
where $\bess$ and $\bess_m$ are the Beta and incomplete Beta functions, respectively  (Abramowitz \&
Stegun 1972)
\begin{eqnarray}
 \bess(a,b)&\equiv &\int_0^1 t^{a-1} (1-t)^{b-1} dt \\
 \bess_m(a,b)&\equiv &\int_0^m t^{a-1} (1-t)^{b-1} dt
\end{eqnarray}
with
\begin{equation}
m= \frac {1+ (\theta/\theta_{c,R})^2}{1+(\Theta/\theta_{c,R})^2}.
\end{equation}
Here
\begin{eqnarray}
S_{X,0} & = & 4.17\times10^{-4} \times (8)^{\alpha_R-1} \left( \frac{E_X}{\mbox{keV}}
\right)^{-\alpha_R} \left( \frac{r_{c,R}} {\mbox{Mpc}}\right) \label{eq.sx0}\\
        & \cdot & \left(\frac{n_{rel,0}}{\mbox{cm}^{-3}\mbox{GeV}^{-1}}\right) \mbox{ erg s}^{-1} \mbox{ cm}^{-2}
        \mbox{ keV}^{-1} \mbox{ arcmin}^{-2},
        \nonumber
\end{eqnarray}
Specifically for Coma, the HXR surface brightness, in the 20--80 keV band, was calculated according to
the choice of parameters described in Sect. 2 and 4, and reported in Tab. \ref{tab.parametri}. To be
noted that $n_{rel,0}$ depends of course only on the value of the central radio emissivity, while
$\langle B \rangle$ represents the volume average out to $R_h=0.9h_{70}^{-1}$ Mpc. The brightness
profiles, as a function of the angular distance from the center, are given in Figs.
\ref{fig.brilhxr2}-\ref{fig.brilhxr05}, which contain, for comparison, the brightness in the same
spectral band of the thermal gas emission. It is evident that, within $\Theta = 32'$ corresponding to
$R_h$, the non-thermal HXR brightness, for appropriate values of $q_e$, can be comparable to, or
greater than the thermal one only for values of $B_0$ less than 0.5 $\mu G$. Beyond $\Theta = 32'$ the
brightness is based upon pure extrapolations outside the maximum extent of the radio-halo, $R_h$, as
measured so far. Spatially resolved HXR observation would shed light on the true distribution of the
relativistic electrons (together with that of the magnetic field) not only within but also outside
$R_h$. It is appealing that the necessary technical capabilities might be available with the next
coming dedicated experiments like, e.g., NeXT (e.g., Takahashi et al. 2004).
\begin{table}[tbp]
\begin{center}
\begin{tabular}{ccccc}
 \hline \hline
 $ q_e $ & $q_b$ & $B_0$     & $<B> $    & $n_{rel,0}$   \\
         &       & $(\mu G)$ & $(\mu G)$ & $(cm^{-3} GeV^{-1})$  \\
 \hline
 0 & 1.87 & 0.1  & 0.03 & $6.46\times10^{-10}$  \\
 2.02 & 1.01 & 0.1 & 0.051 & $6.46\times10^{-10}$ \\
 4.4 & 0 & 0.1 & 0.1 & $6.46\times10^{-10}$ \\
  0 & 1.87 & 0.2  & 0.06 & $1.26\times10^{-10}$ \\
 2.02 & 1.01 & 0.2 & 0.1 & $1.26\times10^{-10}$ \\
 4.4 & 0 & 0.2 & 0.2 & $1.26\times10^{-10}$ \\
  0 & 1.87 & 0.5  & 0.15 & $1.48\times10^{-11}$ \\
 2.02 & 1.01 & 0.5 & 0.255 & $1.48\times10^{-11}$  \\
 4.4 & 0 & 0.5 & 0.5 & $1.48\times10^{-11}$ \\
 0 & 1.87 & 1  & 0.3 & $2.89\times10^{-12}$ \\
 2.02 & 1.01 & 1 & 0.51 & $2.89\times10^{-12}$  \\
 4.4 & 0 & 1 & 1 & $2.89\times10^{-12}$ \\
 0 & 1.87 & 3  & 0.9 & $2.19\times10^{-13}$ \\
 2.02 & 1.01 & 3 & 1.5 & $2.19\times10^{-13}$ \\
 4.4 & 0 & 3 & 3 & $2.19\times10^{-13}$ \\
 0 & 1.87 & 5  & 1.5 & $6.58\times10^{-14}$ \\
 2.02 & 1.01 & 5 & 2.55 & $6.58\times10^{-14}$ \\
 4.4 & 0 & 5 & 5 & $6.58\times10^{-14}$  \\
 \hline
 \end{tabular}
\end{center}
 \caption{\footnotesize{The list of parameters used for the calculation of the Coma HXR brightness.}}
 \label{tab.parametri}
 \end{table}


We wish to conclude this Section with a note on how the volume averaged magnetic field strength should
be computed when using integral values of the radio and the HXR (ICS) flux. Cospatial radio and ICS
flux values can be used to estimate the quantity $(B_{ICS})^{\alpha_R+1} \propto F_{ICS}/F_{radio}$
(e.g. Carilli \& Taylor 2002). In general, except for the case $B$ = constant, the inequality\\
$\langle B^{\alpha_R +1} \rangle \neq  \langle B \rangle^{\alpha_R +1}$.\\
holds, and the use of the right hand side of the inequality to estimate the average value of $B$ (as
in, e.g., Henriksen 1998, Fusco-Femiano et al. 1999, 2004), is formally incorrect, although
quantitatively only by a modest factor. In addition, it must be noted that the value of this quantity
differs from the actual value of $ \langle B \rangle$, which obtains from eq. (\ref{eq.averageb}) and
depends on $q_B$, as it will be shown numerically in the next Section.

\section{Discussion}

This Section will first address the conclusions on $\langle B \rangle$, that follows from available
integral measurements. Then, it will concentrate on the further constraints, that can be derived from
FR measurements.

The integrated HXR flux from Coma has been measured with BeppoSAX and RossiXTE (Fusco-Femiano et al.
1999, 2004, Rephaeli et al. 1999). Both measurements found an excess on top of the thermal emission,
which amounts to $F_{20-80}=(1.5\pm0.5)\cdot10^{-11}$ erg s$^{-1}$ cm$^{-2}$ (Fusco-Femiano et al.
2004), and has been interpreted as ICS emission from the radio--halo. It must be mentioned that the
BeppoSAX result is still controversial (see Rossetti \& Molendi 2004), however, for the purpose of
this discussion, it could at worst be regarded as an upper limit.

\begin{figure}[tbp]
\begin{center}
 \epsfig{file=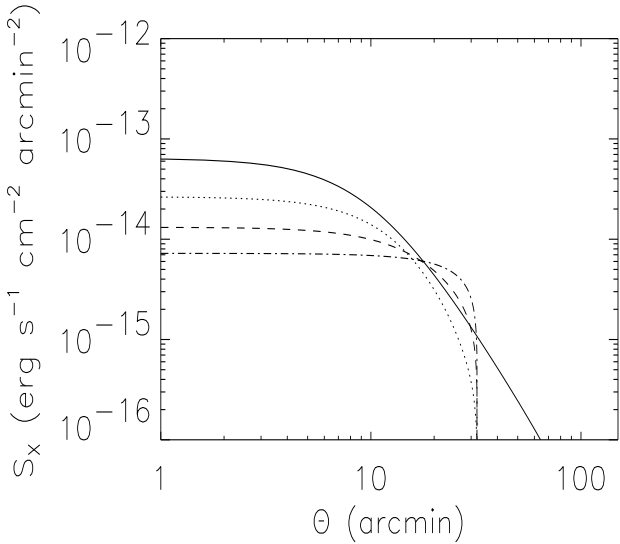,height=8.cm,width=9.cm,angle=0.0}
  \caption{\footnotesize{The 20-80 keV brightness distributions in the Coma cluster,
as produced by ICS of relativistic electrons within a sperical volume corresponding to the radio halo
($\Theta$ = 32 arcmin): the normalizations are chosen to provide an integral value equal to the flux
measured by Fusco-Femiano (2004). The three curves correspond to  $q_e=0$ (dot-dashed), $q_e=2.02$
(dashed) and $q_e=4.4$ (dotted).
 The thermal bremsstrahlung brightness of Coma integrated in the same energy band
  (solid curve) is also shown for comparison.
  }}\label{fig.fusco}
\end{center}
\end{figure}

\begin{figure}[tbp]
\begin{center}
 \epsfig{file=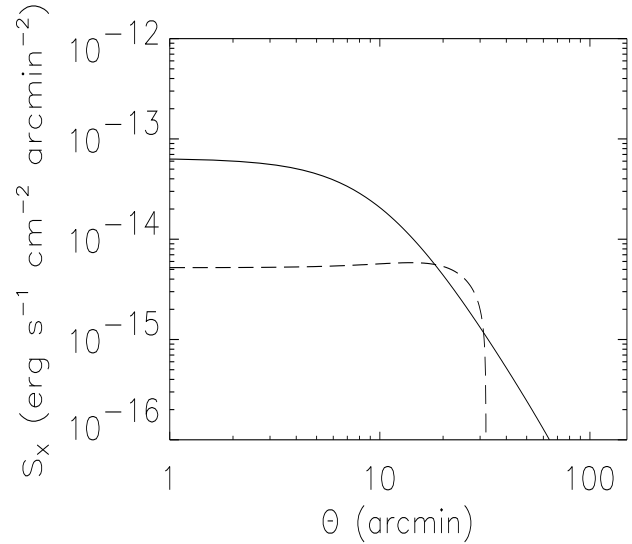,height=8.cm,width=9.cm,angle=0.0}
  \caption{\footnotesize{Same as Fig. \ref{fig.fusco}, for the relativistic
electron distribution given in Fig. \ref{fig.dolag}.
  }}\label{fig.fuscodolag}
\end{center}
\end{figure}

First we consider the case, discussed in Fusco-Femiano et al. (1999, 2004), that this flux is ICS from
within the spherical volume of the radio-halo. Thus we use eq. (\ref{eq.brilhxr}) with $\Theta$ = 32
arcmin, corresponding to $R_h$, and the accordingly different normalization of eq. (\ref{eq.sx0}). By
equating the measured flux to the integral of $S_X$ out to $\Theta$, we obtain the results on $B_0$,
$\langle B \rangle$ and $n_{rel,0}$ given in Table 2, for the three different combinations of $q_e$
and $q_b$, used throughout this paper. The corresponding brightness profiles are shown in Fig.
\ref{fig.fusco}. To be noted that, as anticipated in the previous Section, the three values of
$\langle B \rangle$ are not identical, and that we recover the one obtained by Fusco-Femiano et al.
(2004), namely 0.2 $\mu$G, only if the magnetic field is constant, i.e. $q_B$ = 0.

Within the two extremes for $q_e$, 0 and 4.4, the central value $B_0$ decreases from 0.55 to 0.20
$\mu$G. The intermediate value, $B_0$ = 0.32 $\mu$G is probably the most ``realistic'', in that it
corresponds to a radial decrease in both the relativistic electron density and in the magnetic field
strength.

Furthermore, referring to the considerations devoted, in Sect. 3, to the effects of scalar
fluctuations in the magnetic field, this estimate should, in principle, be considered as an upper
limit: by how much it might differ from reality is hard to tell objectively, it could be evaluated on
the grounds of specific models, a goal beyond the scope of this paper.

On the other hand, some information on the strength of the field, along with the scale of its
vectorial fluctuations, can and have been inferred from Faraday Rotation (FR) measurements, hence we
now turn to their implications.

These measurements have been carried out in several Clusters (see, e.g., Clarke et al. 1999, 2001,
Dolag et al. 2001, 2002) and various papers have been devoted to the uncertainties on the derivation
of values for $B_\parallel$, which are related to the limited statistics on the number of independent
measurements for the same cluster (Newman, Newman \& Rephaeli 2002) or the approximations adopted for
the spectrum of the fluctuations as a function of their scale-length (En\ss lin \& Vogt 2003, Murgia
et al. 2004). When the FR measurements are made using embedded radio sources, the typical central
field strength is found to be $\sim 10-30\mu$G (Eilek 1999), but in general terms FR measurements on
background sources are considered more reliable. For Coma in particular, using data on background
sources, Kim et al. (1991) found a value of $B_\parallel$ for the region within the X-ray core radius
equal to $1.7\pm0.9 \mu$G. Furthermore, Feretti et al. (1995) found that fluctuations probably occur
on all scales down to $\approx 1$ kpc.  For a random vectorial distribution, the previous estimate on
$B_\parallel$ can be translated into an estimate of $B_0$ equal to $3.8\pm2.0 \mu$G.

Taken at face value, this quantity is significantly larger than $B_0$ given in Table 2, even in the
most favourable assumption of $q_e$ = 0. In other words, the interpretation of the excess 20--80 keV
flux, referred to above, in terms of ICS emission from the radio--halo, is hard to reconcile with
these further constraints. The more so if the possible presence of scalar fluctuations, which would
reduce $B_0$ below the strength given in Table 2, is not neglected.

One must however take into account that, for the time being, the excess HXR flux has been detected
only with non-imaging instruments, thus it cannot be excluded that the ICS emission extends outside
$R_h$. The FWHM  of the BeppoSAX PDS instrument used by Fusco-Femiano et al. (2004) covers the Coma
Cluster out to an angular distance from the center $\theta$ = 42 arcmin. If we integrate out to this
radius the predicted ICS profiles, the ones given this time by eq.  (\ref{eq.brilhxr}) and eq.
(\ref{eq.sx0}), we recover the excess HXR flux for values of $B_0$ = 1.37, 0.40, 0.20 $\mu$G for $q_e$
= 0, 2.02 and 4.4 respectively. However, even if we do so, we come close to the central $B$ value
estimated through the FR measurements only with $q_e$ = 0, a rather unlikely hypothesis, although, as
can be judged from Fig. \ref{fig.pressionia}, the non-thermal pressure remains well below the thermal
one out to $R_{vir}$, and therefore cannot be discarded on this ground.

If one accepts this hypothesis, it follows that  most of the HXR emission should be produced by
relativistic electrons located in the outer regions of the cluster. Conversely, the radio-halo
emission is dominated by relativistic electrons located in the central part of Coma and certainly at
distances smaller than $R_h$. The electrons which produce the bulk of the HXR emission are thus
different from those producing the bulk of the radio-halo emission. This is a consequence of the
assumption of an extended, diffuse HXR emission originating from ICS of CMB photons off relativistic
electrons. We note here that the different spatial location of the radio-halo emission and of the bulk
of the HXR emission is also a prediction of various models invoked to reproduce the formation of the
Coma radio-halo (see, e.g., Brunetti et al. 2001, Kuo et al. 2003).

Finally we have considered also the relativistic electrons distribution illustrated in Fig.
\ref{fig.dolag}, which follows from assuming $B(r)$ proportional to $n_{th}(r)$, as from
hydrodynamical simulations (Dolag et al. 2002). The HXR brightness profile is shown in Fig.
\ref{fig.fuscodolag}. Most of the electrons being confined in the outer regions of the radio-halo,
with a density peacking at about 0.8$R_h$, it turns out (upon integration over a spherical volume with
radius $R_h$, last and separate line in Table 2) that $B_0$ = 1.1 $\mu$G, two times larger than for
$q_e$ = 0, and much closer to the estimate from FR measurements. Evidently the same comment applied to
the previous hypothesis remains valid, namely that most of the electrons responsible for the ICS
emission are other than those responsible for most of the radio emission, albeit both are fully
confined within $R_h$.

These considerations make it very clear that in order to properly address the whole issue we badly
need spatially resolved observations in the HXR band, where new generation telescopes are being
designed to operate. Obviously the success will depend more on sensitivity and particle background
than on angular resolution, which could be in the order of one arcmin for this special purpose.

To conclude the discussion, we must remark that the EUV emission excess found in the Coma Cluster, in
the energy band 65-200 eV (see, e.g., Lieu et al. 1996, Bowyer et al. 1999, 2004) does not have any
implication on our results. In fact, relativistic electrons which emit in the mentioned EUV band by
ICS, emit by synchrotron in the frequency range $\nu \sim (0.13-0.4)\,B_\mu$ MHz. Thus, these
electrons should emit synchrotron in the observed radio band ($\nu \geq 30$ MHz) only for magnetic
field $B \geq 75$ $\mu$G, a situation that can be excluded on several grounds, the most obvious being
that the magnetic pressure would largely exceed the thermal pressure. This is not to exclude that the
EUV emission excess might itself be due to IC emission, as proposed for instance by Bowyer et al.
(2004). This is another, open problem, to which we are going to devote an independent paper
(Marchegiani et al., in preparation).

\begin{table}[tbp]
\begin{center}
\begin{tabular}{ccccc}
\hline \hline
 $ q_e $ & $q_b$ & $B_0$         &    $<B>$         & $n_{rel,0}$       \\
         &       & $(\mu G)$     &  $(\mu G)$       & $(cm^{-3} GeV^{-1})$ \\
 \hline
  0      & 1.87  & 0.55          &  0.17        & $1.16\times10^{-11}$  \\
 2.02    & 1.01  & 0.32          &  0.16        & $4.20\times10^{-11}$\\
 4.4    & 0     & 0.20          &  0.20        & $1.29\times10^{-10}$\\
\hline
 - & -          & 1.10          & 0.18         & $2.37\times10^{-12}$\\
 \hline
 \end{tabular}
 \end{center}
 \caption{\footnotesize{The parameters reported in this table are evaluated from the requirement that
 the ICS HXR emission from the radio-halo of Coma, assumed spherical, equals the flux observed by BeppoSAX. We
 report here the central value of the magnetic field $B_0$, the volume-averaged value $<B>$ and the central density
of the relativistic electrons $n_{rel,0}$, for
 the three sets, used throughout this paper, of parameters on the relativistic electron and magnetic
field distributions, which
 reproduce the Coma radio-halo brightness. The separate last line is devoted to the case where the electron distribution
 given in Fig. \ref{fig.dolag} is used.
}} \label{tab.parametri2}
 \end{table}

\section{Summary and conclusions}

Radio halos, that is diffuse synchrotron emission in Clusters of Galaxies, is a rather common
phenomenon which implies the presence of a non-thermal intracluster component, in addition to the
thermal component that gives rise to the diffuse X-ray emission. Unlike the latter, whose density and
temperature distributions can be directly inferred from X-ray measurements, the former is a
combination of relativistic electrons and magnetic fields, whose relative weight in the radio
emissivity cannot be deduced on the basis of the observed radio properties alone. In order to resolve
the degeneracy between these two parameters, an independent estimate of either one or the other is
required.

For the relativistic electrons this estimate can be obtained by measuring their unavoidable emission
through Inverse Compton Scattering on the photons of the Cosmic Microwave Background. It turns out
that, for astrophysically reasonable values of the magnetic field, the ICS emission by the same
electrons responsible for the radio emission should fall in the HXR, roughly in the range 10-100 keV,
and could therefore be easily separated, given adequate sensitivity in that range, from the thermal
emission on spectral grounds.

Alternatively, Faraday Rotation measurements on background radio sources, seen through the cluster,
can provide an absolute estimate of the magnetic field, given the available constraints on the
contribution to the rotation by the thermal electrons.

This paper, using the radio-halo in Coma as a case-study, and its brightness radial distribution as
the only constraint, concentrates on predicting the HXR (20-80 keV) brightness distribution of the ICS
emission, for different, model-independent, central values and radial behaviours of the magnetic field
(or, equivalently, of the density of the relativistic electrons). We had in mind that such predictions
could be valuable to establish requirements on sensitivity for the next generation of hard X-ray
imaging telescopes. Spatially resolved observations would tell us directly the density distribution of
the relativistic electrons, and consequently of the $B$ field. We wish to stress that such
measurements would give us information not only on the smooth radial dependence of $B$, but also of
its scalar fluctuations on small scales. These fluctuations, because of the non-linear dependence of
the radio emissivity on the field strength, imply systematically a smaller number of electrons for the
same radio flux, and consequently they reduce the HXR brightness for the same value of the locally
averaged value of $B$.

The Coma Cluster is a case where detection of a HXR tail on top of the thermal emission has already
been claimed (Fusco-Femiano et al. 2004, Rephaeli et al. 1999). We use the value of this excess flux
to estimate the central value of the field, $B_0$, for three different forms of the radial
distribution of its smooth component, and a random, isotropic distribution of its direction on scales
much smaller than the size of the radio-halo. We do this either assuming that the electrons and
magnetic fields are completely confined within the borders of the radio-halo, as determined so far, or
assuming that the distrutions adopted in our predictions extend out to the virial radius, $R_{vir}$,
of Coma. In the first case we find values of $B_0$ between 0.2 and 0.55 $\mu$G, with the largest value
corresponding to an electron density constant with radius (the volume averaged value of $B$ is exactly
equal to the value of 0.2 $\mu$G obtained by Fusco-Femiano et al. (2004) only if $B$ is assumed
constant). In the second case, taking into account the field of view of the instrument used by
Fusco-Femiano et al. (2004), we find values between 0.2 and 1.4 $\mu$G. Again, the largest value
obtains if the relativistic electron density is assumed constant with radius, a rather unlikely
possibility, which however we cannot exclude simply on dynamical grounds, because their pressure
remains below that of the thermal gas extrapolated out to $R_{vir}$. The intermediate form of the
radial distribution of $B$, yields $B_0$ = 0.4 $\mu$G.

As a third case, we consider the possibility, supported by hydrodynamical simulations (Dolag et al.
2002), that $B(r)$ is proportional to $n_{th}(r)$. This implies a rather peculiar relativistic
electron density distribution, which has its minimum value at the center, and peaks at about 0.8$R_h$.
Upon integration over the spherical volume within $R_h$, the HXR flux measured by Fusco-Femiano et al.
(2004) is recovered with $B_0$ = 1.1 $\mu$G.

A comparison with estimates from FR measurements is then made. Although such estimates are still
controversial, a value of $B_0$ equal to 1-2 $\mu$G seems rather acceptable. Consistency with the
estimates from the HXR excess is at best marginal. The more so if one considers that the latter were
obtained ignoring the (admittedly hard to quantify in a model-independent way) scalar fluctuations,
whose presence would decrease the value of $B_0$. Moreover, if the claimed HXR excess emission is
really there, and it is due to ICS, it would follow that the electrons responsible for the bulk of
this emission are mostly distributed outside the region where the electrons responsible for the bulk
of the radio emission reside. It is then quite evident that measurements of the HXR brightness
distribution, whose prediction is the main scope of this paper, are a necessary step to clarify this
important issue.

\acknowledgements{The authors thank the Referees for useful comments which helped to improve the
paper. S.C. is supported by PRIN-MIUR under contract No.2004027755$\_$003.}

\end{document}